\documentclass[prc,nofootinbib,notitlepage,superscriptaddress]{revtex4-1}

\usepackage{epsfig,amsmath,color,multirow}

\def\snn{\sqrt s_{\rm NN}}

\def\tlc{T_{\rm LCEP}}
\def\tcf{T_{\rm CF}}
\def\gs{\gamma_S}

\newcommand{\be}{\begin{equation}}
\newcommand{\ee}{\end{equation}}                                                                               
\newcommand{\bea}{\begin{eqnarray}}
\newcommand{\eea}{\end{eqnarray}} 

\begin{document}
%*****************************************************************

\title{Hadronization conditions in relativistic nuclear collisions \\ 
and the QCD pseudo-critical line} 

\author{Francesco Becattini}
\affiliation{Universit\`a di Firenze and INFN Sezione di Firenze, Firenze, Italy}

\author{Jan Steinheimer}
\affiliation{Frankfurt Institute for Advanced Studies (FIAS), Frankfurt, Germany}

\author{Reinhard Stock}
\affiliation{Institut f\"ur Kernphysik, Goethe-Universit\"at and 
FIAS, Frankfurt, Germany}

\author{Marcus Bleicher}
\affiliation{Frankfurt Institite for Advanced Studies (FIAS), Frankfurt, Germany}

\begin{abstract}
We compare the reconstructed hadronization conditions in relativistic nuclear collisions
in the nucleon-nucleon centre-of-mass energy range 4.7-2760 GeV in terms of temperature 
and baryon-chemical potential with lattice QCD calculations, by using hadronic multiplicities. 
We obtain hadronization temperatures and baryon chemical potentials with a fit to measured 
multiplicities by correcting for the effect of post-hadronization rescattering. The 
post-hadronization modification factors are calculated by means of a coupled hydrodynamical-transport 
model simulation under the same conditions of approximate isothermal and isochemical 
decoupling as assumed in the statistical hadronization model fits to the data. The 
fit quality is considerably better than without rescattering corrections, as already 
found in previous work.
The curvature of the obtained ``true" hadronization pseudo-critical line $\kappa$ is 
found to be $0.0048\pm0.0026$, in agreement with lattice QCD estimates; the pseudo-critical
temperature at vanishing $\mu_B$ is found to be $164.3\pm1.8$ MeV.
\end{abstract}

\maketitle

%*********************************************************************
\section{Introduction}
%*********************************************************************

It is the goal of Quantum Chromo-Dynamics (QCD) thermodynamics to study the phase 
diagram of strongly interacting matter. Its most prominent feature, the transition 
line between hadrons and partons, in the plane spanned by the baryon-chemical potential 
$\mu_B$ and the temperature $T$, is located in the non-perturbative sector of QCD. 
Here, the theory can be solved on the lattice and has recently led to calculations
of the curvature of the parton-hadron boundary line \cite{kaczm,hotqcd,wup1,wup2,
wup3,pisa1,pisa2,bari}. This line can also be studied experimentally, in relativistic 
collisions of heavy nuclei, where apparently local thermodynamical equilibrium is achieved at 
a temperature well above the (pseudo-)critical QCD temperature $T_c$. Expansion and 
cooling then take the system down to the phase boundary where hadronization occurs. 
We have lately demonstrated \cite{ours1,ours2,ours3} that post-hadronization 
inelastic rescattering, chiefly baryon-antibaryons annihilation, is an important 
feature of the process, which drives the system 
slightly out of equilibrium from the primordial hadronization equilibrium, implying
an actual distinction between hadronization and chemical freeze-out. This rescattering 
stage is taken into account in state-of-the art simulations of the QGP expansion
\cite{song,petersen,karpenko}, where the local equilibrium particle distribution 
(through the so-called Cooper-Frye formula) at some critical values of $T$ and $\mu_B$ 
is used to generate hadrons and resonances which subsequently undergo collisions and decay. 
By calculating the modification of the multiplicities brought about by the rescattering 
stage - the so-called afterburning - it is possible to reconstruct the hadronization 
point by means of a fit to the multiplicities in the framework of the Statistical 
Hadronization Model (SHM). Strictly speaking, this method allows to pin down the 
{\em latest chemical equilibrium point} \cite{ours3} henceforth denoted as LCEP - i.e. 
the point when the primordial chemical equilibrium starts being distorted by the 
afterburning. As equilibrium is an intrinsic feature of hadronization 
\cite{becareview,stock,satz} - as shown by the analysis of elementary collisions - most likely  
LCEP coincides with hadronization itself, as the maintaining of full chemical equilibrium 
in a rapidly expanding hadronic system, for the time needed to produce a measurable 
temperature shift, is highly unlikely. 

As the primordial system temperature (baryon-chemical potential) shifts upward (downward) 
with increasing collision energy, an ascending sequence of experimental energies can, 
thus, map a sequence of LCEPs or hadronization points along the QCD transition line. 
This was the main point of ref.~\cite{ours2} where we showed that, indeed, the reconstructed 
LCEPs seem to follow the extrapolated lattice QCD pseudo-critical line in the $(\mu_B,T)$ 
plane as determined in, e.g., ref.~\cite{wup1}. The agreement between lattice QCD 
calculations and the reconstructed hadronization points in relativistic heavy ion 
collisions seem to imply that, in the examined energy range ($\snn > 7$ GeV), the 
pseudo-critical line has indeed been crossed, and, thus, those energies lie above 
the so-called "onset of deconfinement" \cite{gazdz}. 

This conclusion is less straightforward than it may seem at a glance because, as has
been mentioned, hadron
formation is evidently a universal statistical process \cite{stock,becareview,satz} 
in all kinds of collisions at the same hadronization temperature, with a difference in the 
strangeness sector, whose phase space appears to be only partially filled in elementary 
collisions \cite{becaee,becaee2,becaee3,becapp,satzcast} \footnote {It is worth pointing out here 
that the strangeness undersaturation is still observed in nuclear collisions at high energy 
but it can be accounted for by residual nucleon-nucleon collisions nearby the outer edge of 
the nuclear overlapping region, see \cite{becareview} and references therein.}. 
Indeed, even if the strangeness phase space was fully saturated in elementary collisions, 
if hadron production process in a nuclear collision was fully consistent with 
a picture of subsequent and independent elementary hadronic reactions, strange particle 
production would be strongly suppressed by the exact strangeness conservation over 
the typical small volumes of an elementary collision (canonical suppression) and
subsequent hadronic inelastic collisions would not be able to raise multi-strange particle
abundance to the measured one, as it is shown by transport calculations 
\cite{Petersen:2009zi,Bleicher:2001nz,Drescher:2001hp}. 
Hence, the observation of a strange particle production in agreement with the prediction 
of the SHM for a coherent, large volume, and the agreement between lattice QCD extrapolations
of the pseudo-critical line and the reconstructed hadronization point is a strong
evidence that the pseudo-critical line has been overcome. However, this must cease 
to happen at some sufficiently low centre-of-mass energy, implying the failure of 
at least one of the above conditions. Estimates remain uncertain at present, pointing 
to a region between 4 and 8 GeV.
 
In this paper, we reexamine the hadronization conditions in relativistic heavy ion
collisions over the energy range from low SPS (7.6 GeV) to LHC (2.76 TeV) by using
hadronic multiplicities. We also include the highest AGS energy point at $\snn = 4.5$ 
GeV to probe the aforementioned deconfinement conditions further down in energy. 
For this purpose, we take advantage of an improved initialization of the afterburning 
process by enforcing a particle generation stage - or hydrodynamical decoupling - 
in UrQMD \cite{Bass:1998ca,Bleicher:1999xi,Petersen:2008dd,Huovinen:2012is} at 
fixed values of energy density corresponding to mean temperatures and chemical potentials
equals to those determined in ref.~\cite{ours2} and show how this leads to a further 
remarkable improvement of the fit quality compared to the plain statistical model 
fits \cite{andronic,ours2,ours3}. Finally, we compare the resulting curvature of 
the LCEP-hadronization curve in the $(\mu_B,T)$ plane with the predictions of lattice 
QCD, reporting a good agreement. 

%*********************************************************************
\section{Afterburning and modification factors}
%*********************************************************************

As has been mentioned, we studied the effect of post-hadronization rescattering on 
hadron multiplicities and on the associated SHM fits in previous publications 
\cite{ours1,ours2,ours3} employing a hybrid model \cite{bass,teaney} with a hydrodynamic 
expansion of the QCD plasma terminated at a predefined point where local equilibrium 
particle generation is assumed (hadronization), followed by a hadronic rescattering 
stage modelled by UrQMD \cite{Petersen:2008dd} (afterburning). 
For the fluid dynamical simulation we employed an equation of state which follows from
a so-called "combined hadron-quark model". It is based on a chiral hadronic model which provides
a satisfactory description of nuclear matter properties. The quark phase is introduced 
as a PNJL type model. The transition from the hadronic to the quark phase occurs at about 
$T_c \approx 165$ MeV for $\mu_B=0$, as a smooth crossover as shown in \cite{steinh2}.

For each hadronic species a so-called modification factor is extracted which is defined 
as the ratio between the final multiplicity after the actual chemical (and kinetic) freeze-out 
(which is now species-dependent) and its value without afterburning (at hadronization):
\be
    f_j = \frac{n_j}{n_j^{(0)}}
\ee
The modification factors are then used as additional multiplicative factors to the 
theoretical equilibrium multiplicity yields in the SHM fit, ready to be compared 
to the data. Note that in the calculation of the modification factors, all weak decays 
are turned off, but all strong and EM decays are turned on; this limits the data analysis 
to measurements of multiplicities corrected for the weak decay feed-down.

In our previous studies, the hydrodynamic decoupling procedure was inspired by the so-called 
"inside-outside cascade" mechanism: the transition from the fluid dynamical phase 
to the hadronic transport part occurs in successive transverse slices, of thickness 0.2 fm, 
whenever all fluid cells of that slice fall below a critical energy density, that is six 
times the nuclear ground state density $\epsilon \approx 850 $ $\rm{MeV/fm^3}$. In fact, 
in the present investigation, we have implemented an approximate isothermal termination 
of the hydrodynamical stage at some pre-established temperature $\tcf$ (the subscript CF 
stands for Cooper-Frye). 
This is certainly in much better accordance with the underlying picture of a statistical
hadronization as well as with the previously discussed concept of LCEP, which is 
determined at a fixed value of the proper temperature. For the decoupling, the hypersurface 
is defined by a fixed energy density - at LHC energy - of approximately $0.360 \ 
\mathrm{GeV}/\mathrm{fm}^3$ which corresponds to a mean hadronization temperature 
close to 165 MeV at zero baryon-chemical potential (for the lower energies, see
table~\ref{endenhad}). The cell-to-cell temperature and chemical potential 
fluctuations corresponding to such a hydrodynamic decoupling procedure, at some given
collision energy, are small. For instance, the dispersion of the temperature at LHC 
energy is $\sim 1.5$ MeV, the dispersion of the chemical potential at the SPS energy
is of the order of 10 MeV; these values are comparable or smaller than the parameter
fit errors (see table~\ref{fits}). 
%--------------------------------------------------------------------------------
\begin{table}
\caption{Energy densities used to implement hydrodynamic decoupling or Cooper-Frye
particlization, at the different collision energies. Also quoted are the corresponding
mean temperatures and baryon-chemical potentials.}\label{endenhad}
\vspace{0.5cm}
\begin{tabular}{|c|c|c|c|}
\hline
 $\snn$ (GeV) & Energy density (MeV/fm$^3$) & $\tcf$ (MeV)  & $\mu_{B\, CF}$ (MeV)  \\
\hline    
  4.75        &   508   &  135  &  563      \\          
  7.6         &   435   &  155  &  426      \\
  8.7         &   435   &  161  &  376      \\ 
  17.3        &   435   &  163  &  250      \\
  2760        &   362   &  165  &    0      \\
\hline                
\end{tabular}
\end{table}
%--------------------------------------------------------------------------------

The UrQMD model employs the hypersurface finder outlined in ref.~\cite{Huovinen:2012is}, 
which is used in the Cooper-Frye prescription and sampled to produce hadrons in accordance 
with global conservation of charge strangeness baryon number and the total energy.

It should be pointed out that the calculated modification factors do depend on the 
chosen temperature $\tcf$ ending the hydrodynamical expansion \cite{staibl}. Ideally, 
this should coincide with 
the actual $\tlc$ at each energy, which is {\it a priori} unknown, except for a reasonable 
lower bound set by the chemical freeze-out temperature as determined in the traditional, 
plain, SHM fit. One may then wonder to what extent $\tlc$, which is the outcome of the 
subsequent SHM fit - corrected for afterburning - is affected by the chosen $\tcf$. In 
general, larger $\tcf$ involve larger deviations of the modification factors from unity, 
so one could expect that the fit is influenced to such an extent that the corrected 
SHM fit tends to reproduce the initially chosen value $\tcf$, making the whole method 
non-predictive.
However, this would be the case only if the final particle multiplicities after freeze-out 
were independent of $\tcf$. It was checked that in hybrid simulations this does not happen 
and final particle yields do depend on the chosen decoupling condition. In fact it 
appears that the change in the extracted $\tlc$ tends to be smaller than in the input 
$\tcf$ (i.e. independent of the exact values of the modification), and $\tlc$ shows a 
trend toward a definite value. For instance, for Pb-Pb collisions at $\snn=8.7$ GeV
(see next section for details), for an input $\tcf=144$ MeV we obtained $\tlc=155$ 
MeV whereas for $\tcf=161$ MeV we obtained $\tlc=163$ MeV. In summary, the method
converges. 

The optimal situation, as has been mentioned, is $\tcf=\tlc$, which could be achieved 
by an iterative procedure; however, it would be computationally expensive and not
worth the effort when - in view of the above observation - the difference between 
$\tcf$ and $\tlc$ is only few MeV's. Altogether, the small differences between 
$\tcf$ and $\tlc$ in our analysis (see table~\ref{fits}) in the next section) make 
us confident that the fitted thermal parameters are fully significant, with only a 
marginal dependence on the difference $(\tcf-\tlc)$.

%*********************************************************************
\section{Data analysis}
\label{analysis}
%*********************************************************************

In this section we present the results of our analysis including 5 centre-of-mass
energy points: $\snn = 2.76$ TeV at the LHC, $\snn = 17.3, 8.7, 7.6$ GeV at the SPS
and $\snn = 4.75$ GeV at the AGS.

The modification factors for the strongly stable hadrons have been calculated according 
to the method described in the previous section and are quoted in table~\ref{modf} . 
The decoupling conditions in terms of temperature, that is $\tcf$, and baryon-chemical 
potential are those determined as LCEP's in our previous papers for the most central 
collisions at the LHC \cite{ours3} and for SPS points \cite{ours2} (see table~\ref{fits} 
further below). For the AGS point we did not have any clue about the $\tlc$ value, 
so we iterated the procedure until the fitted $\tlc$ was reasonably close to the 
$\tcf$. The modification factor for the $\phi$ is more difficult to extract than for 
other, strongly stable, particles as it is not the $\phi$ itself which is absorbed, 
but its decay products which rescatter, making $\phi$ reconstruction hardly feasible. 
Assuming that any rescattering of a decay product will lead to a loss of a $\phi$, 
the lower bound of the survival probability has been estimated at 2.76 TeV to be 
about 0.75 \cite{Steinheimer:2012era,Steinheimer:2015msa,Knospe:2015nva}. At all 
lower energies, for afterburning is expected to be less important for this meson, 
we have used an educated guess of 0.875 which is the mean value between 0.75 and 1, 
varying between these bounds in order to check the stability of the best fit solutions.
%--------------------------------------------------------------------------------
\begin{table}
\caption{Afterburning modification factors determined by means of UrQMD (see text 
for explanation). The input decoupling temperatures and chemical potentials are reported in 
table~\ref{fits}}\label{modf}
\vspace{0.5cm}
\begin{tabular}{c|c|c|c|c|c}
\hline
Particle &  $\snn=4.75\;$ GeV  &  $\snn=7.6\;$ GeV & $\snn = 8.7\;$ GeV & $\snn = 17.3\;$ GeV & $\snn = 2760\;$ GeV \\
\hline    
$\pi^+$      &	1.01    & 0.998   & 1.01   &  1.03   & 1.05   \\
$\pi^-$      &	0.980   & 0.980   & 0.997  &  1.03   & 1.05   \\
K$^+$        &	0.947   & 0.927   & 0.918  &  0.928  & 0.918  \\
K$^-$        &	0.890   & 0.901   & 0.868  &  0.891  & 0.919  \\
p            &	0.975   & 0.978   & 0.979  &  0.956  & 0.758  \\
$\bar{\rm p}$&	0.341   & 0.369   & 0.340  &  0.533  & 0.754  \\
$\Lambda$    &	0.981   & 0.935   & 0.932  &  0.927  & 0.816  \\
$\bar\Lambda$&	0.432   & 0.478   & 0.422  &  0.601  & 0.821  \\
$\Xi^-$      &	1.01    & 0.979   & 0.977  &  0.970  & 0.886  \\
$\bar\Xi^+$  &	0.573   & 0.575   & 0.531  &  0.698  & 0.876  \\
$\Omega$     &	0.888   & 0.882   & 0.808  &  0.873  & 0.789  \\
$\bar\Omega$ &	0.596   & 0.573   & 0.566  &  0.706  & 0.778  \\
$\phi$       &     -    &    -    &   -    &  -      & 0.75   \\
\hline
\hline                
\end{tabular}
\end{table}
%--------------------------------------------------------------------------------

The particle set used in the analysis is the intersection between the available
measured multiplicities and the set of particles for which a modification factor 
was calculated, see table~\ref{dataset}. All data refer to central collisions of 
Au+Au (at the AGS) and Pb+Pb (at SPS and LHC). As has been mentioned, we have confined 
ourselves to data sets where weak feed-down was subtracted in order to make a proper 
comparison between corrected (with modification factors) and non-corrected fits.
%--------------------------------------------------------------------------------
\begin{table}
\caption{Particle multiplicities measured at various centre-of-mass energies employed in our analysis. 
The data are $4\pi$ multiplicities except at $\snn= 2760$ GeV where they are midrapidity densities.}
\label{dataset}
\vspace{0.5cm}
\begin{tabular}{c|c|c|c|c|c}
\hline
Particle &  $\snn=4.75\;$ GeV  &  $\snn=7.6\;$ GeV & $\snn = 8.7\;$ GeV & $\snn = 17.3\;$ GeV & $\snn = 2760\;$ GeV \\
\hline    
$\pi^+$      &	133.7$\pm$9.9 \cite{ags2}  & 241$\pm$12 \cite{na49}    & 293$\pm$15.3 \cite{na49}  &  619$\pm$35.4 \cite{na49}    & 733$\pm$54 \cite{alice1}   \\
$\pi^-$      &	    -                      & 274$\pm$14 \cite{na49}    & 322$\pm$16.3 \cite{na49}  &  639$\pm$35.4 \cite{na49}    & 732$\pm$52 \cite{alice1}   \\
K$^+$        &	23.7$\pm$2.86 \cite{ags1}  & 52.9$\pm$3.6 \cite{na49}  & 59.1$\pm$3.55 \cite{na49} &  103$\pm$7.1  \cite{na49}    & 109$\pm$9  \cite{alice1}   \\
K$^-$        &	3.76$\pm$0.47 \cite{ags1}  & 16$\pm$0.45 \cite{na49}   & 19.2$\pm$1.12 \cite{na49} &  51.9$\pm$3.55 \cite{na49}   & 109$\pm$9  \cite{alice1}   \\
p            &1.23$\pm$0.13$^a$\cite{ags5} &  -                        &  -                        &  -                           & 34$\pm$3   \cite{alice1}   \\
$\bar{\rm p}$&     -                       & 0.26$\pm 0.04^b$           &  -                          &  4.23$\pm 0.35^c$        & 33$\pm$3  \cite{alice1}   \\
$\Lambda$    &	18.1$\pm$1.9   \cite{ags3} & 36.9$\pm$3.3  \cite{na49}  & 43.1$\pm$4.32 \cite{na49}  &  48.5$\pm$8.6 \cite{na49}  & 26.1$\pm$2.8 \cite{alice2} \\
$\bar\Lambda$&	0.017$\pm$0.005 \cite{ags4}& 0.39$\pm$0.045 \cite{na49} & 0.68$\pm$0.076 \cite{na49} &  3.32$\pm$0.34 \cite{na49} & -              \\
$\Xi^-$      &	 -                         & 2.42$\pm$0.345 \cite{na49} & 2.96$\pm$0.04 \cite{na49}  &  4.40$\pm$0.64 \cite{na49} & 3.57$\pm 0.27^c$ \cite{alice3} \\
$\bar\Xi^+$  &	 -                         & 0.120$\pm$0.036 \cite{na49}& 0.13$\pm$0.022 \cite{na49} &  0.71$\pm$0.1  \cite{na49} & 3.47$\pm 0.26^c$ \cite{alice3} \\
$\Omega$     &	 -                         &     -                      &0.14$\pm 0.05^{b}$\cite{na49}&  0.59$\pm$0.11 \cite{na49}& 1.26$\pm 0.22^{d,e}$ \cite{alice3} \\
$\bar\Omega$ &	 -                         &     -                      &   -                        &0.260$\pm$0.067 \cite{na49} &  -            \\
$\phi$       &   -                         & 1.84$\pm$0.36 \cite{na49}  & 2.55$\pm$0.25\cite{na49}   &  8.46$\pm$0.50 \cite{na49} & 13.8$\pm$1.77 \cite{alice4} \\
$B^{f}$      &  363$\pm$10  \cite{ags1}    & 349$\pm$5.1   \cite{na49}  &  349$\pm$5.1 \cite{na49}   &  362$\pm$8 \cite{na49}     &  -             \\ 
\hline
\multicolumn{6}{l}{$^a$ This is the ratio p/$\pi^+$}\\
\multicolumn{6}{l}{$^b$ Our extrapolation \cite{ours1} based on measurements in ref.~\cite{alt}} \\
\multicolumn{6}{l}{$^c$ Our extrapolation \cite{ours1} of spectra measured in ref.~\cite{anticic}. The NA49 data
compilation \cite{na49} quotes 4.25$\pm$0.28 by M. Utvic.} \\
\multicolumn{6}{l}{$^d$ $\Omega+\bar\Omega$} \\
\multicolumn{6}{l}{$^e$ Interpolation to 0-5\% centrality quoted in ref.~\cite{ours3}} \\
\multicolumn{6}{l}{$^f$ Number of participants = net baryon number of the fireball} \\
\hline                
\end{tabular}
\end{table}
%--------------------------------------------------------------------------------

\begin{figure}[ht]
\begin{center}
\includegraphics[width=0.6\columnwidth]{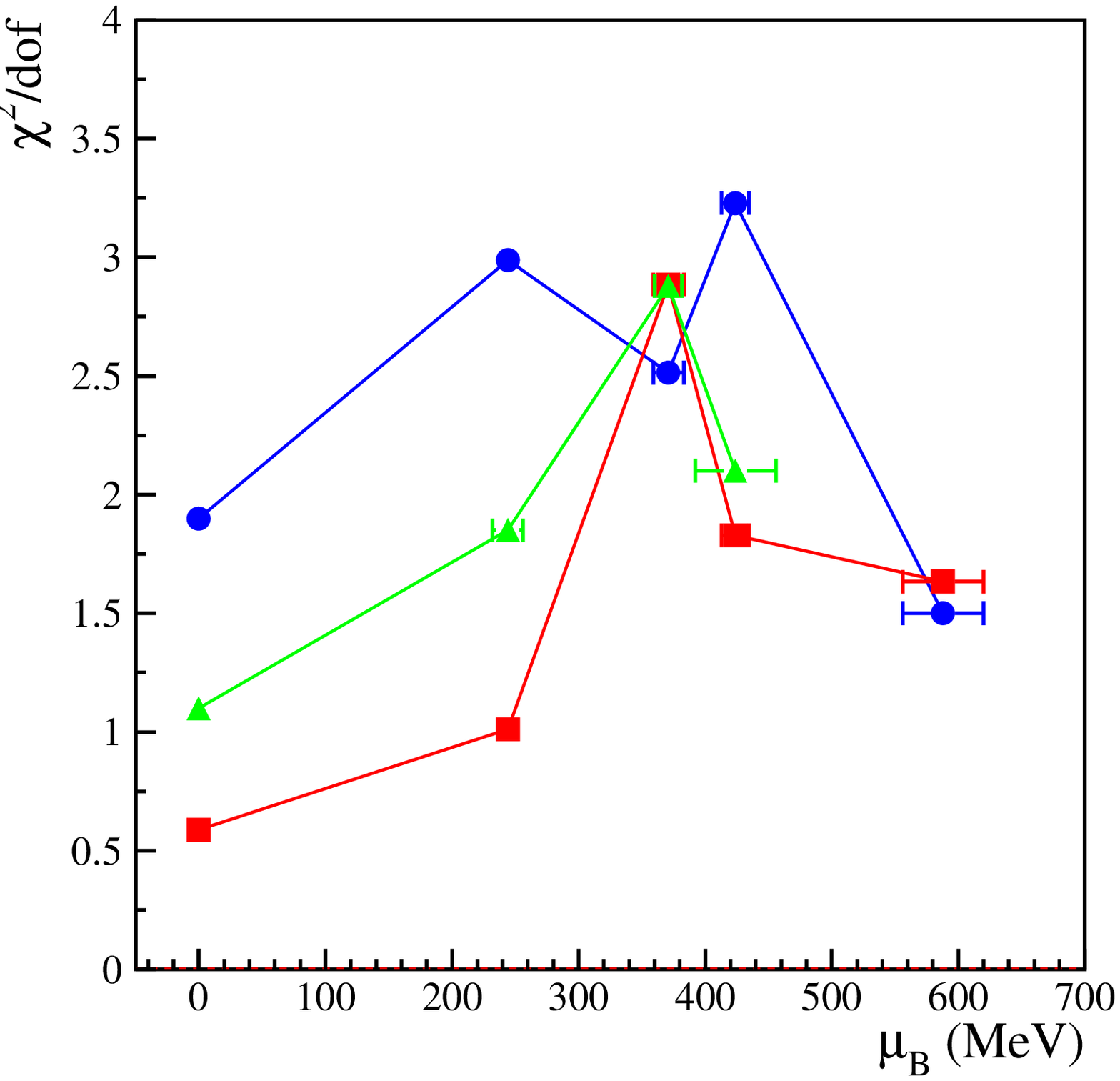}
\caption{(Color online) $\chi^2/dof$ of the SHM multiplicity fits at the 5 different
energies. Blue dots: plain SHM fit without afterburning corrections. Green dots:
SHM fit with afterburning correction with the isochronous decoupling method (points
from refs.~\cite{ours2}. Red dots: SHM fits with afterburning corrections with the 
new, approximately isothermal decoupling.}
\label{chi2}
\end{center}
\end{figure}

The SHM, the formulas for primary and final multiplicities, the fitting procedure
with and without modification factors have been described in detail elsewhere \cite{ours1}.
Herein, we simply summarize the obtained results in table~\ref{fits}. The corrected 
fits are of remarkable better quality with respect to the plain SHM fits, as it is
shown in fig.~\ref{chi2}, confirming previous findings. One exception stands out, the
SPS point at 8.7 GeV, which is the point where the ratio K$^+/\pi^+$ attains its maximum
observed value \cite{horn}. Indeed, the measured ratio K$^+/\pi^+$ overshoots 
the statistical model prediction \cite{bgkms} by more than 2$\sigma$ (see table~\ref{kpi}), a 
discrepancy which is not cured by the afterburning correction.
%--------------------------------------------------------------------------------
\begin{table}
\caption{Measured vs fitted K$^+$ multiplicities at $\snn=7.6$ and 8.7 GeV, in the
so-called horn region, with corresponding deviations in units of $\sigma$ within 
round brackets.}\label{kpi}
\vspace{0.5cm}
\begin{tabular}{|c|c|c|c|c|}
\hline
   $\snn$ (GeV)  &  Measured     &  Plain fit    &  Modified fit  \\   
\hline 
   7.6           & 52.9$\pm$3.6  &   47.1 (-1.6) &  45.3 (-2.1)   \\
   8.7           & 59.1$\pm$3.6  &   51.4 (-2.2) &  49.6 (-2.7)   \\
\hline      
\end{tabular}
\end{table}
%--------------------------------------------------------------------------------

Notably, the $\chi^2/dof$ is of full statistical significance
once the modification factors are introduced in the two highest energy points. Moreover,
there is a further improvement of the fit quality with respect to the previous, 
non-isothermal, fits \cite{ours2,ours3}.
%--------------------------------------------------------------------------------
\begin{table}
\caption{Results of the SHM fits with and without afterburning corrections. In the last
column we quote the corresponding decoupling temperatures $\tcf$ and chemical potentials
$\mu_{B\, {\rm CF}}$ employed to calculate the modification factors}
\label{fits}
\vspace{0.5cm}
\begin{tabular}{|c|c|c|c|}
\hline
 Parameters     & Without afterbuner      & With afterburner  & CF in URQMD   \\        
\hline
\multicolumn{4}{|c|}{Au-Au $\snn=4.75$ GeV} \\
\hline               
$T$ (MeV)       & 122.1$\pm$4.0   & 130.5$\pm$12.3  &  135       \\
$\mu_B$ (MeV)   & 563$\pm$15      & 588$\pm$32      &  563       \\	   
$\gs$           & 0.638$\pm$0.074 & 0.71$\pm$0.12  &  1.0       \\
$\chi^2$/dof    & 4.5/3           & 4.9/3           &    \\              
\hline
\multicolumn{4}{|c|}{Pb-Pb $\snn=7.6$ GeV} \\
\hline               
$T$ (MeV)       & 139.6$\pm$3.7   & 157.7$\pm$4.3  &   155    \\
$\mu_B$ (MeV)   & 437$\pm$20      & 424$\pm$11     &   426    \\	   
$\gs$           & 0.922$\pm$0.075 & 0.871$\pm$0.059&   1.0    \\
$\chi^2$/dof    & 22.6/7          & 12.8/7         &          \\
\hline
\multicolumn{4}{|c|}{Pb-Pb $\snn=8.7$ GeV} \\
\hline               
$T$ (MeV)       & 148.2$\pm$3.8   & 163.3$\pm$5.0   &   161    \\
$\mu_B$ (MeV)   & 385$\pm$11      & 371$\pm$12      &   376    \\	   
$\gs$           & 0.783$\pm$0.062 & 0.773$\pm$0.055 &   1.0    \\
$\chi^2$/dof    & 17.6/7          & 20.2/7          &          \\
\hline
\multicolumn{4}{|c|}{Pb-Pb $\snn=17.3$ GeV} \\
\hline                
$T$ (MeV)       & 150.4$\pm$3.9    & 162.3$\pm$2.7   &   163    \\
$\mu_B$ (MeV)   & 265$\pm$10       & 244$\pm$6       &   250    \\	   
$\gs$           & 0.914$\pm$0.052  & 0.885$\pm$0.029 &   1.0    \\
$\chi^2$/dof    & 26.9/9           & 9.1/9           &          \\
\hline
\multicolumn{4}{|c|}{Pb-Pb $\snn=2760$ GeV} \\
\hline                
$T$ (MeV)       & 155.0$\pm$3.7 & 163.8$\pm$3.3 &  165    \\
$\mu_B$ (MeV)   & 0 (fixed)     & 0 (fixed)     &  0      \\	   
$\gs$           & 1.07$\pm$0.05 & 1.02$\pm$0.04 &  1.0    \\
$\chi^2$/dof    & 15.2/8        & 4.7/8         &         \\
\hline                
\end{tabular}
\end{table}
%--------------------------------------------------------------------------------
%----------------------------------------------------------------------------------
%----------------------------------------------------------------------------------

The quoted errors in table~\ref{fits} are the fit errors. There are additional small
systematic uncertainties on the fit parameters related to the errors on the modification
factors. These errors stem from the uncertainties on the cross-sections used in 
UrQMD and from finite Monte Carlo statistics. The former are difficult to estimate, 
whereas the latter are simpler; in our runs they are of the order of few percent 
for all particles, thus they do not imply any significant variation of the best
fit parameter values. The only largely unknown modification factor is, as has been 
mentioned, the $\phi$'s, for which we could obtain an estimate of 0.75 at the top energy
point $\snn=2760$ GeV. As the rescattering of a neutral meson is expected to diminish
in a lower multiplicity environment, one can reasonably set a lower bound of 0.75 
at all lower energies. To estimate the effect of the uncertainty, we have varied
the $\phi$ modification factor to 0.75 and 1 in turn at each energy point. The resulting
variation of the fit parameters is within 1 MeV for the temperature and few MeV's
for the baryon chemical potentials so that the relative systematic error is always
less than 1\%, thus below the fit error.

%*********************************************************************
\section{Curvature of the pseudo-critical line}
\label{curvature}
%*********************************************************************

Finally, we have used the LCEP points in table~\ref{fits} to fit the curvature of 
the pseudo-critical line in the $(\mu_B,T)$ plane. The curvature parameter $\kappa$
is defined by the equation:
\be\label{formula}
T_c(\mu_B) = T_c (0) \left[ 1-\kappa \left( \frac{\mu_B}{T_c(0)} \right)^2 \right] 
\ee
which is the same formula used in lattice calculations. As the LCEP points in $(\mu_B,T)$ 
plane have errors on both coordinates, we have minimized the $\chi^2$:
\be
  \chi^2 = \sum_{i=1}^N (Z_i - Z_{0i})^T C_i^{-1} (Z_i-Z_{0i})
\ee
where 
$$
Z_i = (\mu_{Bi},T_i) \qquad Z_{0i} = (\mu_{Bi}^0,T_c(\mu_{Bi}^0))
$$
In the above equation, $\mu_{Bi}$ and $T_i$ are the output of the SHM fit for
the $i$th centre-of-mass energy point, while 
$C_i$ is their corresponding covariance matrix. The $\mu_{Bi}^0$'s are free parameters 
which represent the ``true" values of the chemical potential in the fitted curve,
$T_c(\mu_{Bi}^0)$ being the corresponding ``true" temperatures.
Therefore, the free parameters in this fit are the chemical potentials $\mu_{Bi}^0$, 
whose position is strongly constrained by the "measured" values $\mu_{Bi}$, the 
value of the pseudo-critical temperature $T_c(0)$ and $\kappa$.
%--------------------------------------------------------------------------------
\begin{table}
\caption{Best fit parameters and $\chi^2$ values for the fit to the reconstructed
LCEPs and chemical freeze-out points in the $(\mu_B,T)$ plane.}\label{criticline}
\vspace{0.5cm}
\begin{tabular}{|c|c|c|c|c|}
\hline
 Fit method         & $T_c(0)$ (MeV)  & $\kappa$          &  $\lambda$        & $\chi^2$    \\   
\hline
  4 points            & 164.3$\pm$1.8 & 0.0048$\pm$0.0026 &  -                  & 0.47/2    \\ 
  4 points, freezeout & 157.4$\pm$6.2 & 0.013$\pm$0.0072  &  -                  & 1.0/2     \\
  5 points            & 167.7$\pm$4.0 & 0.0111$\pm$0.0055 &  -                  & 2.8/3     \\ 
  5 points, freezeout & 162.1$\pm$4.4 & 0.020$\pm$0.004   & -                   & 2./3      \\
 Quartic 5 points     & 164.4$\pm$2.7 & 0$\pm$0.0091      &  0.0109$\pm$0.00047 & 0.97/2    \\ 
\hline      
\end{tabular}
\end{table}
%--------------------------------------------------------------------------------

\begin{figure}[ht]
\begin{center}
\includegraphics[width=0.6\columnwidth]{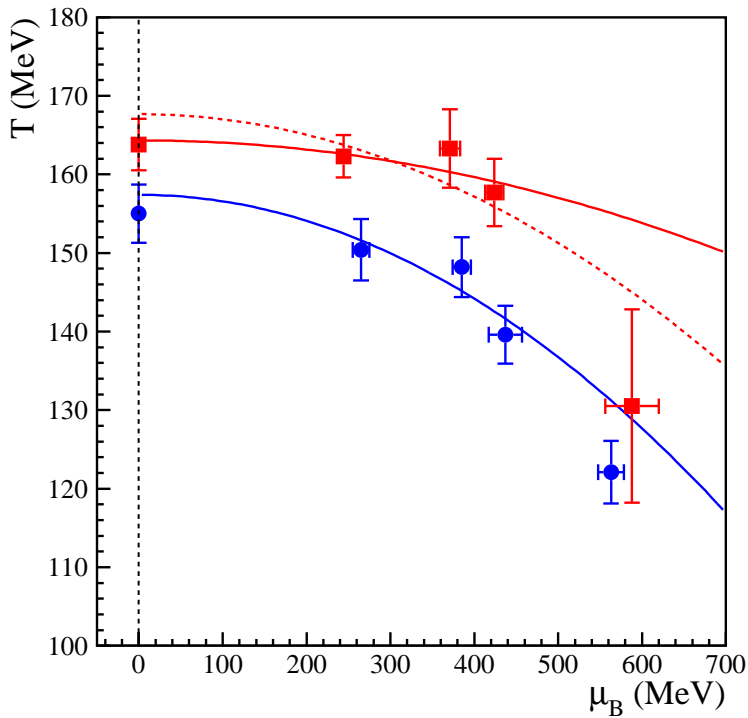}
\caption{(Color online) Reconstructed LCEPs (red squared dots) vs plain chemical freeze-out
fitted points (blue round dots) in the $(\mu_B,T)$ plane. The solid lines are the 4 points
quadratic fits quoted in table~\ref{criticline}; the dashed line is the 5 point quadratic
fit including the lowest energy AGS point.}
\label{tmu}
\end{center}
\end{figure}

It should be pointed out that the equation (\ref{formula}) is a quadratic approximation 
of the actual pseudo-critical line, hence deviations are expected at large values 
of the chemical potentials. Therefore, we have first excluded the lowest energy 
point and made a fit to the four highest energy points at our disposal. We have also 
compared with the freezeout points, for which many systematic studies have been done in
the past \cite{cleymans}. The fitted values of $T_c(0)$ and $\kappa$ are reported in 
table~\ref{criticline} while the fitted curves are shown in fig.~\ref{tmu}. It can 
be seen that the fit quality is excellent for the LCEP points and satisfactory for 
the plain freeze-out points. The systematic error on the curvature due to the uncertainty 
in the $\phi$ meson modification factors has been obtained repeating the fit with the 
varied $(\mu_B,T)$ points and turned out to be $0.0006$.
%----------------------------------------------------------------------------------
%----------------------------------------------------------------------------------

The lowest energy point falls below the fit curve in both cases. There are three
possible explanations for this:
\begin{enumerate}
\item{} the (mundane) effect of having excluded it from the fit;
\item{} the quadratic approximation in (\ref{formula}) falls short at such large
chemical potential values;
\item{} the lowest energy point did not reach the pseudo-critical transition line,
and so the onset of deconfinement can be located between 4.5 and 7.6 GeV.
\end{enumerate}
The latter hypothesis is indeed the most intriguing, but its very consideration requires
the ruling out the first two. 
If the AGS point is included, the fit quality is not as good as the 4 point fit, 
still it is within statistical significance, as it can be seen in table~\ref{criticline}.
Yet, there is some tension between the fitted curve and the two extremal points 
(LHC and AGS) which both undershoot the curve by 4 and 15 MeV respectively, as it
can be seen in fig.~\ref{tmu}). On the other hand, including a quartic term $\lambda 
(\mu_B/T_c(0))^4$ improves the fit but the limited number of points and the limited 
range does not allow to pin down both the quadratic and the quartic term at the same
time; indeed, the fit has multiple solutions and the best fit is awkwardly found for 
$\kappa \simeq 0$ (see table~\ref{criticline}). 
%--------------------------------------------------------------------------------
\begin{table}
\caption{Comparison between the curvature $\kappa$ in lattice QCD calculations
and our estimate.}\label{lqcd}
\vspace{0.5cm}
\begin{tabular}{|c|c|}
\hline
   Reference             & $\kappa$            \\   
\hline
  This work - 4 points   & 0.0048$\pm$0.0026   \\
  This work - 5 points   & 0.0111$\pm$0.0055   \\
   \cite{kaczm}          & 0.0066$\pm$0.0005   \\ 
   \cite{hotqcd}         & $(0.0033 - 0.0123)$ \\ 
   \cite{wup2} - Max     & 0.020$\pm$0.002     \\ 
   \cite{wup2} - Min     & 0.0066$\pm$0.00020  \\
   \cite{wup3}           & 0.0149$\pm$0.0021   \\
   \cite{pisa2}          & 0.0135$\pm$0.0020   \\
   \cite{bari}           & 0.020$\pm$0.004     \\
\hline      
\end{tabular}
\end{table}
%--------------------------------------------------------------------------------

Finally, returning to fig.~\ref{tmu}, which is our main result showing the estimated
QCD pseudo-critical curve in the $(\mu_B,T)$ plane, it is appropriate to compare it 
with recent lattice QCD calculations (see table~\ref{lqcd}). Because of the pseudo-critical 
nature of the transition, both $T_c(0)$ and $\kappa$ parameters depend on the observable 
used to define it \cite{wup2}. It could be therefore expected that these parameters
will somewhat differ from those extracted with the fluctuation of conserved charges 
\cite{fluct}, which can be directly calculated in lattice QCD but are definitely 
less robust observables in relativistic heavy ion collisions with respect to mean 
multiplicities \cite{koch}.

In our comparison, we have quoted all recent literature on the 
subject. We find that our main value of $0.0048$ is in slightly better agreement with 
lower estimates \cite{wup2,kaczm,hotqcd}. We also note that our value is compatible
with a recent estimate based on a comparison between lattice QCD and data \cite{bazavov}.

%*********************************************************************
\section{Conclusions}
\label{conclu}
%*********************************************************************

In summary, we have determined the hadronization conditions (strictly speaking the 
latest chemical equilibrium points) in relativistic heavy ion collisions, by using an
improved calculation of the post-hadronization rescattering correction. The quality
of the statistical model fit considerably improves with respect to traditional fits
without afterburning corrections as well as with respect to our previous calculations.
The pseudo-critical temperature at zero $\mu_B$, determined with hadronic multiplicities,
turns out to be $T_c = 164$ MeV, which is significantly higher than lattice QCD calculations
based on different observables. We find good agreement between the extracted curvature 
of the hadronization curve and the corresponding QCD lattice calculations for the 
pseudo-critical line, with a preference for the lower estimates.

At this stage, it is not possible to make a definite statement about the crossing 
of the pseudo-critical line at the lowest energy point at $\snn=4.7$  GeV. This is 
due to lack of appropriate data and this issue will be tackled by future experiment 
in that energy range (NA61 at SPS and the facilities NICA and FAIR). Our observations 
might have interesting implications for the location of the critical point 
\cite{stephanov}; we note that a recent analysis \cite{lacey} locate it at $T \simeq 165$ 
MeV and $\mu_B \simeq 95$ MeV which just sits on our hadronization curve.

%************************************************************
\section*{Acknowledgments}
%************************************************************

We are grateful to C.~Bonati and V.~Koch for useful suggestions.\\ 
This work was supported by GSI and BMBF and by the University 
of Florence grant {\it Fisica dei plasmi relativistici: teoria
e applicazioni moderne}. The computational resources were provided 
by the LOEWE Frankfurt Center for Scientific Computing (LOEWE-CSC).


\begin{thebibliography}{99}
%************************************************************
\section*{References}
%************************************************************


\bibitem{kaczm}
 O.~Kaczmarek {\it et al.}, Phys.\ Rev.\ D {\bf 83} (2011) 014504.

\bibitem{hotqcd} 
  P.~Hegde {\it et al.} [Bielefeld-BNL-CCNU Collaboration],
  arXiv:1511.03378 [hep-lat].

\bibitem{wup1}
 G.~Endrodi, Z.~Fodor, S.~D.~Katz and K.~K.~Szabo, JHEP {\bf 1104} (2011) 001.

\bibitem{wup2}
 S.~Borsanyi, G.~Endrodi, Z.~Fodor, S.~D.~Katz, S.~Krieg, C.~Ratti and K.~K.~Szabo,
  JHEP {\bf 1208} (2012) 053.
  
\bibitem{wup3}
  R.~Bellwied, S.~Borsanyi, Z.~Fodor, J.~Günther, S.~D.~Katz, C.~Ratti and K.~K.~Szabo,
  Phys.\ Lett.\ B {\bf 751} (2015) 559.
  
\bibitem{pisa1}
 C.~Bonati, M.~D'Elia, M.~Mariti, M.~Mesiti, F.~Negro and F.~Sanfilippo,
 Phys.\ Rev.\ D {\bf 90} (2014) no.11,  114025.

\bibitem{pisa2}
  C.~Bonati, M.~D'Elia, M.~Mariti, M.~Mesiti, F.~Negro and F.~Sanfilippo,
  Phys.\ Rev.\ D {\bf 92} (2015) no.5,  054503.

\bibitem{bari}
  P.~Cea, L.~Cosmai and A.~Papa, Phys.\ Rev.\ D {\bf 93} (2016),  014507.
  
\bibitem{ours1}  
  F.~Becattini, M.~Bleicher, T.~Kollegger, M.~Mitrovski, T.~Schuster and R.~Stock,
  Phys.\ Rev.\ C {\bf 85} (2012) 044921.
   
\bibitem{ours2}
  F.~Becattini, M.~Bleicher, T.~Kollegger, T.~Schuster, J.~Steinheimer and R.~Stock,
  Phys.\ Rev.\ Lett.\  {\bf 111} (2013) 082302.

\bibitem{ours3}
 F.~Becattini, E.~Grossi, M.~Bleicher, J.~Steinheimer and R.~Stock,
  Phys.\ Rev.\ C {\bf 90} (2014) no.5,  054907.
  
\bibitem{song}
  X.~Zhu, F.~Meng, H.~Song and Y.~X.~Liu, Phys.\ Rev.\ C {\bf 91} (2015) no.3,  034904.
 
\bibitem{petersen} 
  J.~Auvinen and H.~Petersen, Phys.\ Rev.\ C {\bf 88} (2013) no.6, 064908.
 
\bibitem{karpenko}
  I.~Karpenko, M.~Bleicher, P.~Huovinen and H.~Petersen, arXiv:1601.00800.

\bibitem{becareview}
  F.~Becattini, {\em An Introduction to the Statistical Hadronization Model}
  arXiv:0901.3643; F.~Becattini and R.~Fries, Landolt-Bornstein {\bf 23} (2010) 208
  [arXiv:0907.1031].
  
\bibitem{stock}
   R.~Stock, Phys.\ Lett.\ B {\bf 456} (1999) 277.

\bibitem{satz}
  H.~Satz, Eur.\ Phys.\ J.\ ST {\bf 155} (2008) 167.

\bibitem{gazdz} 
 M.~Gazdzicki, M.~Gorenstein and P.~Seyboth, Acta Phys.\ Polon.\ B {\bf 42} (2011) 307 
   
\bibitem{becaee}
  F.~Becattini, Z.\ Phys.\ C {\bf 69} (1996) no.3,  485; F.~Becattini, A.~Giovannini and S.~Lupia,
  Z.\ Phys.\ C {\bf 72} (1996) 491.
  
\bibitem{becaee2}  
 F.~Becattini, P.~Castorina, J.~Manninen and H.~Satz, Eur.\ Phys.\ J.\ C {\bf 56} (2008) 493.  
 
\bibitem{becaee3}
 L.~Ferroni and F.~Becattini, Eur.\ Phys.\ J.\ C {\bf 71} (2011) 1824. 

\bibitem{becapp}
  F.~Becattini and G.~Passaleva, Eur.\ Phys.\ J.\ C {\bf 23} (2002) 551.

\bibitem{satzcast}
  P.~Castorina and H.~Satz, Adv.\ High Energy Phys.\  {\bf 2014} (2014) 376982.

\bibitem{Petersen:2009zi} 
  H.~Petersen, M.~Mitrovski, T.~Schuster and M.~Bleicher,
  Phys.\ Rev.\ C {\bf 80}, 054910 (2009).

\bibitem{Bleicher:2001nz} 
  M.~Bleicher, F.~M.~Liu, A.~Keranen, J.~Aichelin, S.~A.~Bass, F.~Becattini, 
  K.~Redlich and K.~Werner,  Phys.\ Rev.\ Lett.\  {\bf 88}, 202501 (2002)
  
\bibitem{Drescher:2001hp} 
  H.~J.~Drescher, J.~Aichelin and K.~Werner,
  Phys.\ Rev.\ D {\bf 65}, 057501 (2002) 

\bibitem{Bass:1998ca} 
  S.~A.~Bass {\it et al.},
  Prog.\ Part.\ Nucl.\ Phys.\  {\bf 41}, 255 (1998)

\bibitem{Bleicher:1999xi} 
  M.~Bleicher {\it et al.},
  J.\ Phys.\ G {\bf 25}, 1859 (1999)

\bibitem{Petersen:2008dd} 
  H.~Petersen, J.~Steinheimer, G.~Burau, M.~Bleicher and H.~St\"ocker, Phys.\ Rev.\ C {\bf 78}, 044901 (2008) 

\bibitem{Huovinen:2012is} 
  P.~Huovinen and H.~Petersen, Eur.\ Phys.\ J.\ A {\bf 48}, 171 (2012).

\bibitem{andronic}
 J.~Stachel, A.~Andronic, P.~Braun-Munzinger and K.~Redlich, J.\ Phys.\ Conf.\ Ser.\  {\bf 509} (2014) 012019.

\bibitem{steinh2} 
  J.~Steinheimer, S.~Schramm and H.~St\"ocker, Phys.\ Rev.\ C {\bf 84}, 045208 (2011).

\bibitem{staibl}
  J.~Steinheimer, J.~Aichelin and M.~Bleicher, Phys.\ Rev.\ Lett.\  {\bf 110} (2013) 042501.

\bibitem{bass}
  S.~A.~Bass and A.~Dumitru, Phys.\ Rev.\ C {\bf 61} (2000) 064909.

\bibitem{teaney}
  D.~Teaney, J.~Lauret and E.~V.~Shuryak, Nucl.\ Phys.\ A {\bf 698} (2002) 479.

\bibitem{ags2} 
  L. Ahle {\it et al.} (E-802 Collaboration), Phys. Rev. C {\bf 59} (1991) 2173.

\bibitem{ags1} 
  L. Ahle {\it et al.} (E-802 Collaboration), Phys. Rev. C {\bf 60} (1999) 044904. 

\bibitem{ags5} 
  L. Ahle {\it et al.} (E802 Collaboration), Phys. Rev. C {\bf 60} (1999) 064901.

\bibitem{ags3} 
  S.~Ahmad {\it et al.}, Phys. Lett. B {\bf 382} (1996) 35.

\bibitem{ags4} 
  S. Albergo {\it et al.}, Phys. Rev. Lett. {\bf 88} (2002) 062301.

\bibitem{Steinheimer:2012era} 
  J.~Steinheimer, J.~Aichelin and M.~Bleicher,
  EPJ Web Conf.\  {\bf 36}, 00002 (2012). 
  
\bibitem{Steinheimer:2015msa} 
  J.~Steinheimer and M.~Bleicher, EPJ Web Conf.\  {\bf 97}, 00026 (2015).  
 
\bibitem{Knospe:2015nva} 
  A.~G.~Knospe, C.~Markert, K.~Werner, J.~Steinheimer and M.~Bleicher,
  Phys.\ Rev.\ C {\bf 93}, no. 1, 014911 (2016).

\bibitem{na49} 
 NA49 collaboration compilation of numerical results
 https://edms.cern.ch/ui/file/1075059/4/na49\_compil\_20130801.pdf and references therein.

\bibitem{alice1}
  B.~Abelev {\it et al.} [ALICE Collaboration], Phys.\ Rev.\ C {\bf 88} (2013) 044910.
 
\bibitem{alice2}
  B.~B.~Abelev {\it et al.} [ALICE Collaboration], Phys.\ Rev.\ Lett.\  {\bf 111} (2013) 222301.

\bibitem{alice3}  
 B.~B.~Abelev {\it et al.} [ALICE Collaboration], Phys.\ Lett.\ B {\bf 728} (2014) 216;
   Erratum: Phys.\ Lett.\ B {\bf 734} (2014) 409.

\bibitem{alice4}
 B.~B.~Abelev {\it et al.} [ALICE Collaboration], Phys.\ Rev.\ C {\bf 91} (2015) 024609.

\bibitem{alt}
  C. Alt {\it et al.} [NA49 Collaboration], Phys. Rev. C {\bf 73} (2006) 044910.

\bibitem{anticic}
 T.~Anticic {\it et al.} [NA49 Collaboration], Phys.\ Rev.\ C {\bf 83} (2011) 014901.

\bibitem{horn}
  C.~Alt {\it et al.} [NA49 Collaboration], Phys.\ Rev.\ C {\bf 77} (2008) 024903.

\bibitem{bgkms}
  F.~Becattini, M.~Gazdzicki, A.~Keranen, J.~Manninen and R.~Stock, Phys.\ Rev.\ C 
 {\bf 69} (2004) 024905.  

\bibitem{cleymans}
  J.~Cleymans, H.~Oeschler, K.~Redlich and S.~Wheaton, Phys.\ Rev.\ C {\bf 73} (2006) 034905.

\bibitem{fluct} F.~Karsch, Central Eur. J. Phys. {\bf 10} (2012) 1234; A. Bazavov
{\it et al.}, Phys. Rev. Lett. {\bf 109} (2012) 192302; P. Alba, W. Alberico, 
R. Bellwied, M. Bluhm, V. Mantovani Sarti, M. Nahrgang, and C. Ratti, Phys. Lett. B 
{\bf 738} (2014) 305; S. Borsanyi, Z. Fodor, S. D. Katz, S. Krieg, C. Ratti, and K. K.
Szabo, Phys. Rev. Lett. {\bf 113} (2014) 052301.

\bibitem{koch}
  P.~Braun-Munzinger, V.~Koch, T.~Schäfer and J.~Stachel, Phys.\ Rept.\  {\bf 621} (2016) 76.

\bibitem{bazavov}
  A.~Bazavov {\it et al.}, Phys.\ Rev.\ D {\bf 93} (2016),  014512.
 
\bibitem{stephanov}
  M.~A.~Stephanov, Prog.\ Theor.\ Phys.\ Suppl.\  {\bf 153} (2004) 139
   [Int.\ J.\ Mod.\ Phys.\ A {\bf 20} (2005) 4387].

\bibitem{lacey}
  R.~A.~Lacey, Phys.\ Rev.\ Lett.\  {\bf 114} (2015) no.14, 142301.

\end{thebibliography}
\end{document}